\documentclass[]{spie}  
\usepackage[normalem]{ulem}
\usepackage{multirow}

\usepackage{amsmath,amsfonts,amssymb}
\usepackage{graphicx}
\usepackage[colorlinks=true, allcolors=blue]{hyperref}

\title{A VLBI receiving system for the South Pole Telescope}

\author[a]{Junhan Kim}
\author[a]{Daniel P. Marrone}
\author[b]{Christopher Beaudoin}
\author[c,d,e,f]{John E. Carlstrom}
\author[g]{\mbox{Sheperd S. Doeleman}}
\author[a]{Thomas W. Folkers}
\author[a]{David Forbes}
\author[a]{Christopher H. Greer}
\author[a]{\mbox{Eugene F. Lauria}}
\author[a]{Kyle D. Massingill}
\author[c]{Evan Mayer}
\author[a,h]{Chi H. Nguyen}
\author[a]{George Reiland}
\author[b]{Jason SooHoo}
\author[g]{Antony A. Stark}
\author[g]{Laura Vertatschitsch}
\author[g]{Jonathan Weintroub}
\author[g]{\mbox{Andr\'{e} Young}}

\affil[a]{Department of Astronomy and Steward Observatory, University of Arizona, 933 N. Cherry Avenue, Tucson, AZ 85721, USA}
\affil[b]{Massachusetts Institute of Technology, Haystack Observatory, Route 40, Westford, MA 01886, USA}
\affil[c]{Kavli Institute for Cosmological Physics, University of Chicago, 5640 South Ellis Avenue, Chicago, IL 60637, USA}
\affil[d]{Department of Astronomy and Astrophysics, University of Chicago, 5640 South Ellis Avenue, Chicago, IL 60637, USA}
\affil[e]{Department of Physics, University of Chicago, 5640 South Ellis Avenue, Chicago, IL 60637, USA}
\affil[f]{Enrico Fermi Institute, University of Chicago, 5640 South Ellis Avenue, Chicago, IL 60637, USA}
\affil[g]{Harvard-Smithsonian Center for Astrophysics, 60 Garden Street, Cambridge, MA 02138, USA}
\affil[h]{Center for Detectors, School of Physics and Astronomy, Rochester Institute of Technology, 1 Lomb Memorial Dr., Rochester, NY 14623, USA}

\authorinfo{Further author information: (Send correspondence to Junhan Kim)\\Junhan Kim: E-mail: junhankim@email.arizona.edu, Telephone: 1 520 621 2288}

\begin{document} 
\maketitle

\begin{abstract}
The Event Horizon Telescope (EHT) is a very-long-baseline interferometry (VLBI) experiment that aims to observe supermassive black holes with an angular resolution that is comparable to the event horizon scale. The South Pole occupies an important position in the array, greatly increasing its north-south extent and therefore its resolution.

The South Pole Telescope (SPT) is a 10-meter diameter, millimeter-wavelength telescope equipped for bolometric observations of the cosmic microwave background. To enable VLBI observations with the SPT we have constructed a coherent signal chain suitable for the South Pole environment. The dual-frequency receiver incorporates state-of-the-art SIS mixers and is installed in the SPT receiver cabin. The VLBI signal chain also includes a recording system and reference frequency generator tied to a hydrogen maser. Here we describe the SPT VLBI system design in detail and present both the lab measurements and on-sky results.
\end{abstract}

\keywords{EHT, SIS receiver, SPT, submillimeter, VLBI}

\section{INTRODUCTION}
\label{sec:intro}  

The Event Horizon Telescope\footnote{\indent http://www.eventhorizontelescope.org/} (EHT) is a very-long-baseline interferometry (VLBI) experiment operating at observing frequencies of 230 and 345~GHz\cite{2009astro2010S..68D}. The EHT aims to study the immediate environment of supermassive black holes such as Sagittarius A* (Sgr~A*) at the center of our galaxy\cite{2008Natur.455...78D} and the black hole in the center of galaxy M87\cite{2012Sci...338..355D} with angular resolution sufficient to resolve the event horizons of these black holes. The EHT array is composed of submillimeter telescopes and telescope arrays around the globe, each of which is outfitted with precise time standards and systems for fast digitization and recording of data. As of 2017, the EHT has performed its first 230~GHz observation with an array consisting of the Submillimeter Array (SMA) and the James Clerk Maxwell Telescope (JCMT) in Hawaii, the Submillimeter Telescope (SMT) in Arizona, the Large Millimeter Telescope (LMT) in Mexico, the Institute de Radioastronomie Millim\'{e}trique (IRAM) 30 m telescope at Pico Veleta, Spain, the Atacama Large Millimeter/submillimeter Array (ALMA)\cite{2018PASP..130a5002M} and the Atacama Pathfinder Experiment (APEX)\cite{2015A&A...581A..32W} in Chile, and the South Pole Telescope (SPT) in Antarctica.

The inclusion of the SPT is crucial to the EHT because of its geographic location. The South Pole is not only an outstanding place for submillimeter observations, due to its low precipitable water vapor and stable atmosphere\cite{2016PASP..128g5001R}, but it also provides the most extended baselines in the array when the SPT is paired with the other EHT sites, most of which are located in the northern hemisphere. For example, the baseline between the SPT and the SMT is greater than 10,000~km and gives 15~micro-arcsecond ($\sim$ 14 G$\lambda$) angular resolution at 345~GHz. For the main target, Sgr~A*, the apparent diameter of the black hole event horizon is approximately 50~$\mu$as, and this high resolution is necessary for imaging of the innermost region around the black hole event horizon. Polarization-sensitive EHT observations also enable the study of magnetic structure in the accretion flow\cite{2015Sci...350.1242J}. Finally, the SPT can serve as a continuous observing partner for any other EHT station because Sgr~A* never sets at the South Pole. This allows the longest time series observations, which will be an important resource for time variability studies of this unique object.

The SPT is a 10-meter diameter, off-axis telescope built to observe the cosmic microwave background (CMB) radiation at millimeter wavelengths\cite{2011PASP..123..568C}. It is sited at the Dark Sector Lab (DSL) of the Amundsen-Scott South Pole Station, Antarctica, along with the Background Imaging of Cosmic Extragalactic Polarization (BICEP) project. The SPT CMB camera has gone through several upgrades, including adding polarization sensitivity and increasing the number of detectors in the array\cite{2009AIPC.1185..475C, Austermann:2012jl, Benson:2014br}. Currently, the third generation SPT-3G is in operation. Since the camera uses a transition-edge sensor (TES) bolometer array that is insensitive to the phase of incoming radiation, a new coherent signal chain is required to perform interferometric observation in coordination with the other EHT sites. We have developed a dual-frequency VLBI receiver system to incorporate the SPT to the EHT array. Both 230 and 345 GHz receivers facilitate dual-polarization, two-single-sideband observations. We have deployed the receiver system including hydrogen maser and the VLBI recording setup to the South Pole. The 230 GHz receiver had the first on-sky test in January 2015 and successfully detected an interferometric fringe with the APEX\cite{Kim2018CenA}. It began scientific operation with the EHT observation in April 2017. In this paper, we present the receiver design, optics, and software, and report the lab test results of the system components.

\section{RECEIVER SYSTEM}
\label{sec:rxsystem}

\begin{figure}[t]
   \begin{center}
   \begin{tabular}{c}
   \includegraphics[height=12cm]{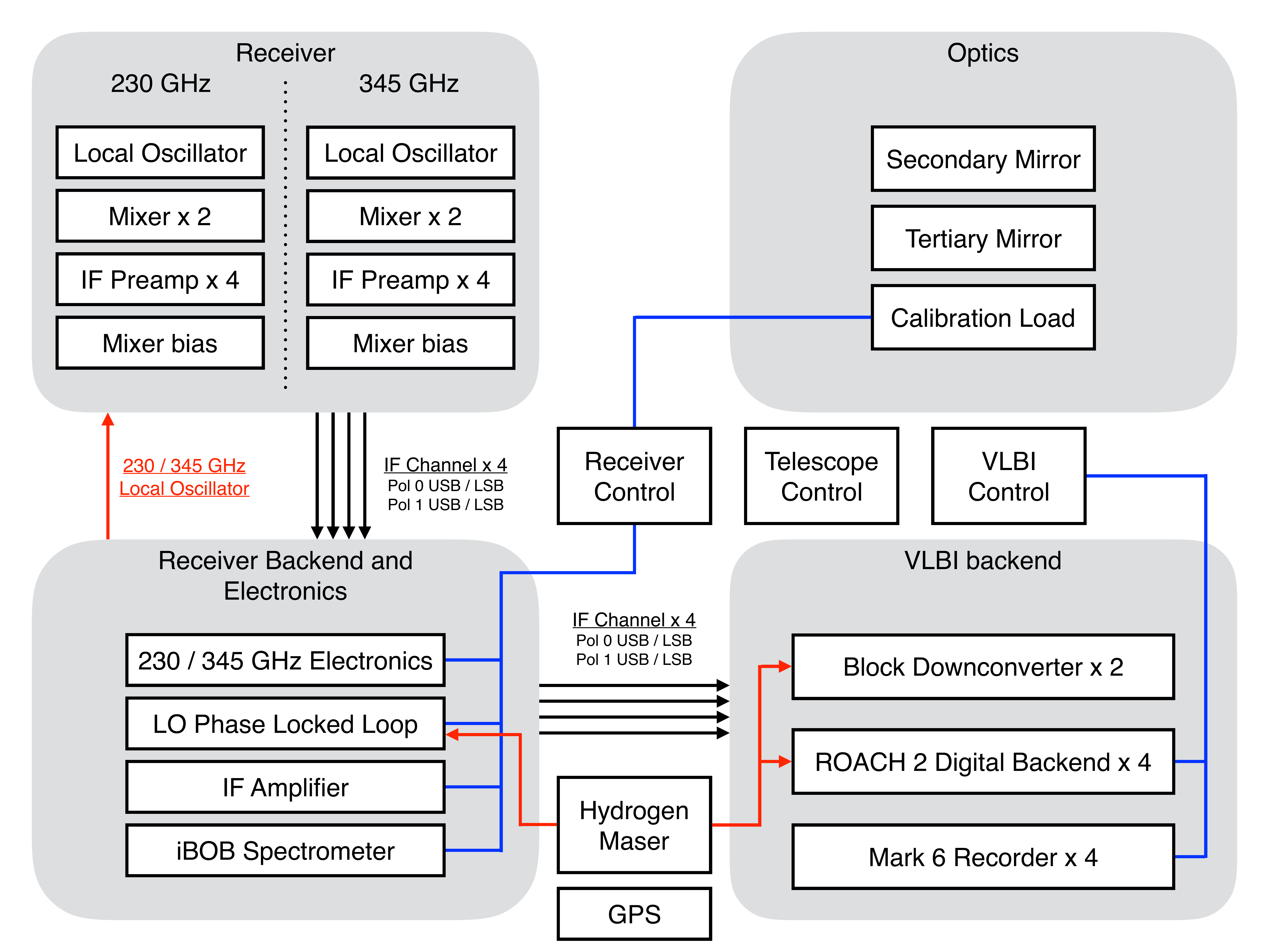}
   \end{tabular}
   \end{center}
   \caption[]
   {\label{fig:system} SPT VLBI receiver system diagram. The coherent receiver works at 230 and 345 GHz frequencies. Receiver electronics provide LO locked to hydrogen maser to the mixers as well as deliver the IF signal to the VLBI backend for digitization and recording. Receiver control, telescope control, and VLBI control computers run software for the system.}
   \end{figure}

The VLBI receiving system of the SPT comprises the receiver, the receiver electronics, VLBI backend, and optics as illustrated schematically in Figure~\ref{fig:system}. The receiver combines electromagnetic waves from the sky with a high-purity and stable reference tone (local oscillator; LO) in a superconducting mixer, downconverting the sky signal to an intermediate frequency (IF) of a few GHz. The IF signal is amplified through the receiver electronics and forwarded to the VLBI backend. The VLBI backend digitizes the analog signal and records the data to arrays of hard disk drives with accurate timestamping. The whole receiver system is synchronized with the 10 MHz reference signal from the hydrogen maser, an atomic clock. In this section, we explain each of the receiver system element. We will describe the optics separately in the next section.

\subsection{Receiver}
\label{sec:rxsystem_rx}

The receiver operates at both 230 and 345 GHz frequency bands. Following ALMA terminology, we will often refer to the 230 GHz and 345 GHz portions of the receiver as band 6 (211$-$275 GHz) and band 7 (275$-$373 GHz), respectively. Telescopes in the EHT use identical LO and IF frequencies, with the IF chosen to match the ALMA receivers. The LO frequencies were chosen to avoid the rest frequencies of carbon monoxide (CO), which would absorb emission from the galactic center, and maximize atmospheric transmission in both sidebands in bands 6 and 7. Table~\ref{tab:eht_freq} shows the LO frequency, IF and the corresponding sky frequency of each band. The band 7 mixer for the SPT receiver is under development as of May 2018. 
\begin{table}[t]
\vspace{10pt}
\caption{SPT VLBI receiver frequency setup} 
\label{tab:eht_freq}
\begin{center}
\begin{tabular}{|l|l|l|}
\hline
\rule[-1ex]{0pt}{3.5ex}  & Band 6 & Band 7 \\
\hline
\rule[-1ex]{0pt}{3.5ex}  Gunn oscillator frequency & 73.7 GHz & 114.2 GHz \\
\hline
\rule[-1ex]{0pt}{3.5ex}  Local oscillator (LO) frequency & 221.1 GHz & 342.6 GHz \\
\hline
\rule[-1ex]{0pt}{3.5ex}  Intermediate frequency (IF) & 5$-$9 GHz & 4$-$8 GHz \\
\hline
\multirow{2}{*}{\hspace{0.75ex}Sky frequency}& 212.1$-$216.1 GHz & 334.6$-$338.6 GHz \\
& 226.1$-$230.1 GHz & 346.6$-$350.6 GHz \\
\hline
\end{tabular}
\end{center}
\end{table}

\begin{figure}[t]
   \begin{center}
   \begin{tabular}{c}
   \includegraphics[height=10cm]{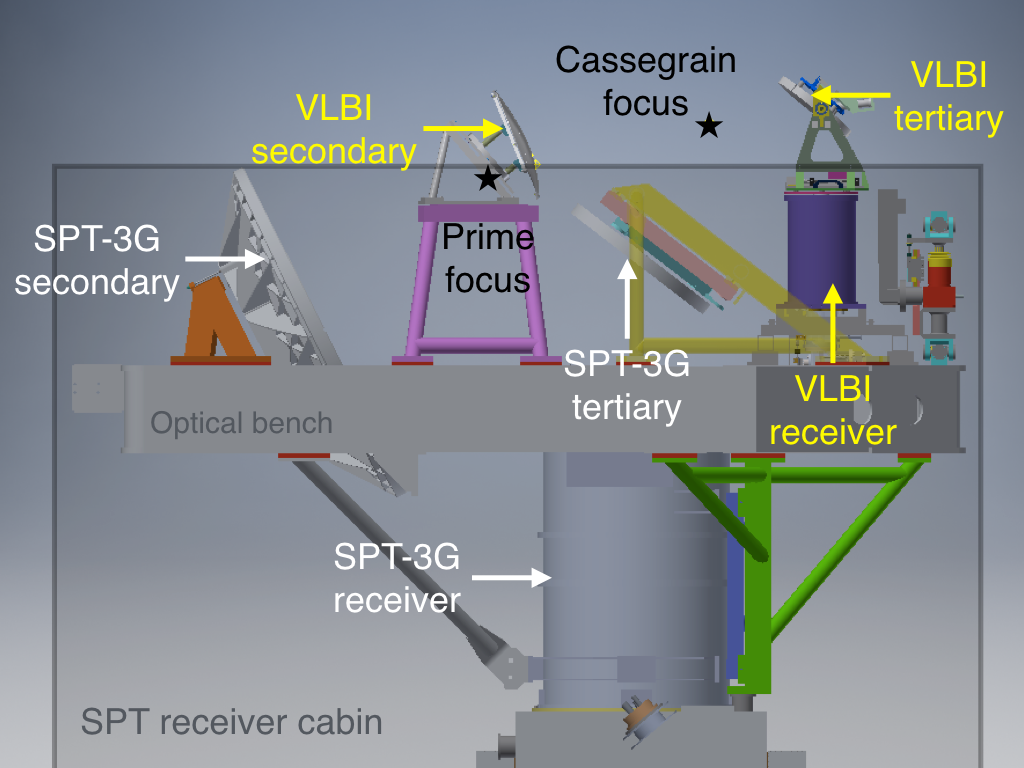}
   \end{tabular}
   \end{center}
   \caption[]
   {\label{fig:optics_model} CAD model of the SPT receiver cabin. The SPT-3G and VLBI systems are indicated by white and yellow arrows, respectively. Prime and Cassegrain foci are shown with black stars. The primary is located beyond the right side of the figure. The grey box shows the location of receiver cabin roof and walls.}
   \end{figure}

\begin{figure}[ht]
   \begin{center}
   \begin{tabular}{cc}
   \includegraphics[height=7.5cm]{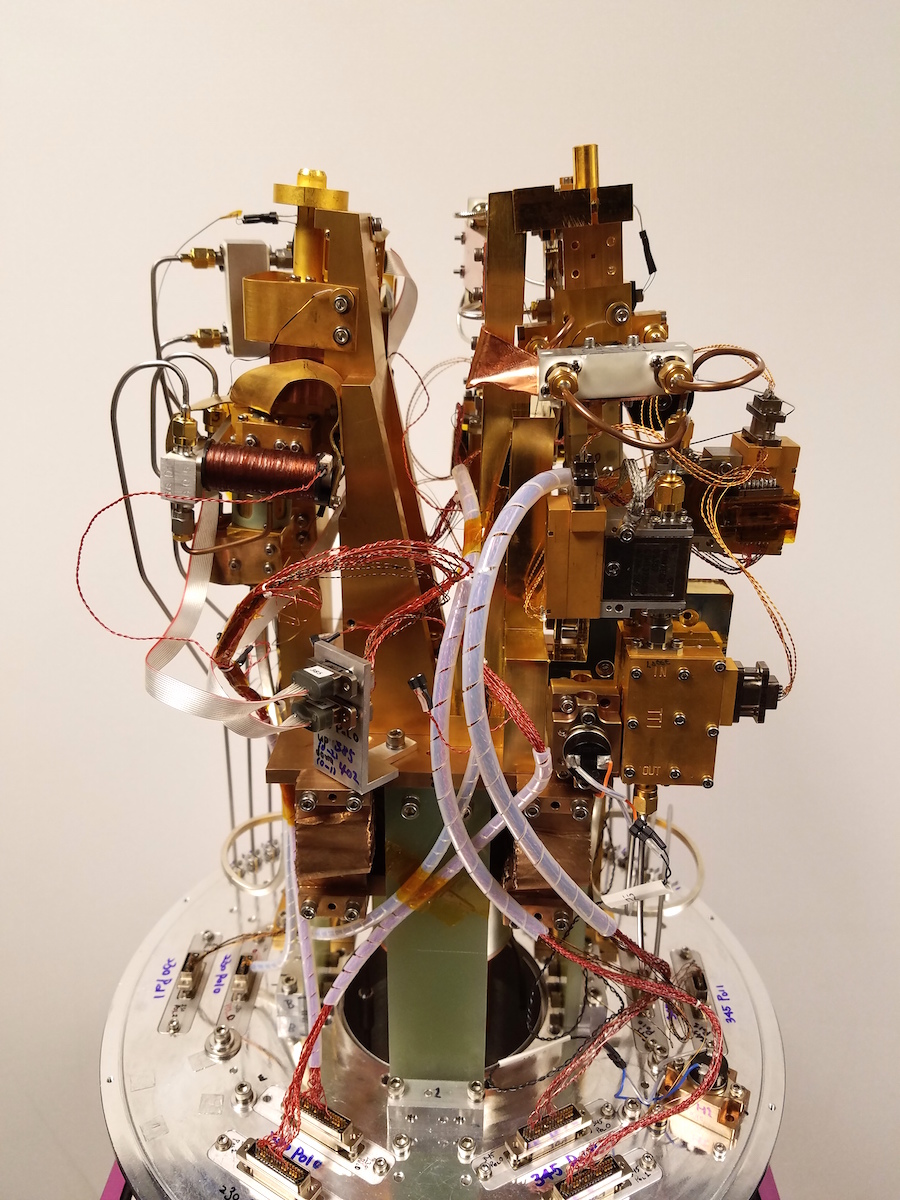} & \includegraphics[height=7.5cm]{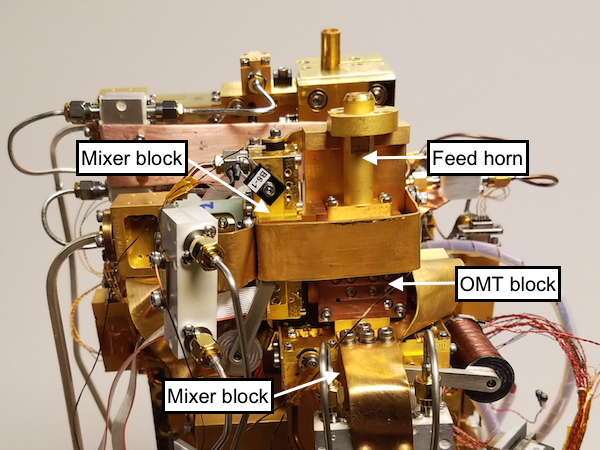}
   \end{tabular}
   \end{center}
   \caption[] 
   { \label{fig:rx_assy} {\it Left}: The 4~K stage of the receiver. The left half is the 230 GHz receiving system, the right half is the 345 GHz system. Feed horns for both frequencies are tilted by 5.74 degrees with respect to the central vertical plane to share the rotatable tertiary mirror. {\it Right}: 230 GHz receiver assembly. Feed horn, OMT, and mixer blocks are attached to the second refrigeration stage and cooled to 4~K.}
   \end{figure}

The SPT VLBI receiver incorporates bands 6 and 7 in a package that fits within the confined space of the climate-controlled SPT receiver cabin. The cabin is dominated by the SPT-3G receiver, its optics, and optical path, so the VLBI receiver is positioned behind the SPT-3G tertiary mirror towards the primary (Figure~\ref{fig:optics_model}) and illuminated by a separate optical system. 
The receiver cryostat surrounds a Sumitomo RDK-408D2P closed-cycle refrigerator, which is connected to an F-70L helium compressor. The two-stage Gifford-McMahon cold head achieves temperatures of 43~K for the first and 4~K for the second stages. Eight LakeShore DT-670 silicon diodes monitor both stages of the cold head and the mixer block temperatures.
The feed horns, ortho-mode transducers (OMTs), and mixers are cooled to 4~K, and the rest of the cold assembly including frequency triplers for the LOs, LO waveguide, and wiring harnesses are coupled to the first stage (Figure~\ref{fig:rx_assy}). Gold plated heat straps between the mixer blocks and the second stage ensure that the blocks cool efficiently. To prevent the thermal contact between the stages via the electrically conductive LO waveguide, we use waveguide thermal isolators with periodic bandgap structure\cite{2003stt..conf..148H} supported by G-10 fiberglass. There are two sets of isolators, separating the ambient and first-stage waveguide segments, and the first and the second stage segments. 

\begin{figure}[ht]
   \vspace{10pt}
   \begin{center}
   \begin{tabular}{cc}
   \includegraphics[height=7.5cm]{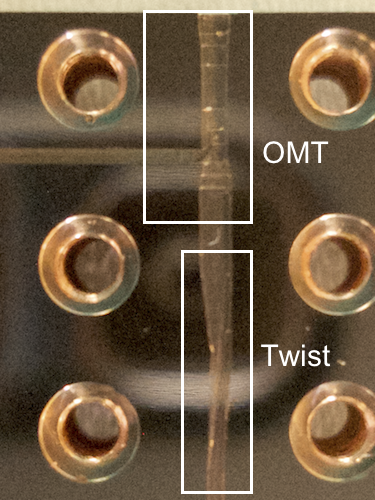} & \includegraphics[height=7.5cm]{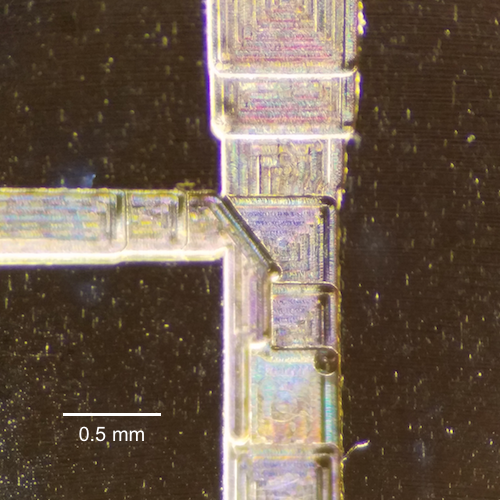}
   \end{tabular}
   \end{center}
   \caption[] 
   { \label{fig:omt} {\it Left}: The inner surface of the split block OMT. We have a polarization twist on one end of the OMT. {\it Right}: The T-shaped waveguide of the OMT for the polarization separation of the input signal.}
   \end{figure}

The 230 GHz receiver employs an ALMA band 6 corrugated feed horn and two mixer/preamplifier modules developed by the National Radio Astronomy Observatory (NRAO)\cite{2014ITTST...4..201K}. The mixers are driven by the LO signal generated from a Spacek Labs bias-tuned Gunn oscillator and tripled by Virginia Diodes WR3.4$\times$3 broadband triplers. The LO is waveguide-injected. The fixed-frequency Gunn oscillator reduces complexity compared to frequency/backshort tuning for broadband oscillators, which is useful for winter operation of the receiver. We fabricated a waveguide quadrature hybrid\cite{SIRKANTH:gZT_O6qs} to equally divide the Gunn output for the two mixer blocks, and electronically control the LO power to each using QuinStar Technology PIN diode variable attenuators. The LO system, including a harmonic mixer and a cross-guide coupler, is attached outside the dewar, and waveguide vacuum feedthroughs\cite{ediss:2005vr} bridge the vacuum shield. All the waveguide designs of the assembly follow the ALMA standard\cite{Kerr99waveguideflanges}. 
Each mixer block delivers two isolated sidebands in a single polarization, with a 4-12 GHz IF. The polarization splitting is achieved by the OMT, which is a version of 
the design in Ref.~\citenum{2009stt..conf..191D}, proportionally scaled to operate in 230 GHz frequency band. The S-parameters and the polarization isolation of the scaled design were simulated using the frequency domain solver of the CST Microwave studio. Figure~\ref{fig:omt} shows the cross-section of the OMT block. The OMT separates the input radiation from the horn to two linearly polarized signals. We add a polarization twist \cite{Chattopadhyay:2010kba} at the end of one OMT output simplify the geometry of the mixer blocks in the dewar. The IF output of the mixers are sent via stainless coaxial cable to the SMA feedthroughs on the bottom of the dewar, and then to the electronics rack inside the cabin through LMR-240 coaxial cables.

The receiver assembly is surrounded by an aluminum radiation shield with Mylar super-insulation. There are separate vacuum windows above the two feed horns. The size of the window is chosen such that it is greater than five times the beam waist size. We use Z-cut quartz manufactured by Boston Piezo-Optics for the windows and bonded Teflon sheets to both sides of the window to implement anti-reflection (AR) coating\cite{2001stt..conf..410K}. We estimate the insertion loss to be less than 0.05 dB in the receiver sky frequency bands, from the thickness measurement of the Teflon glued window. An AR-coated quarter-wave plate is installed on top of the window to convert circular polarization to linear polarization. We also ran mechanical finite element analysis (FEA) of the dewar. For dewar tilt angles between 0 and 90 degrees, the maximum displacement of the horn aperture is less than 80 $\mu$m, corresponding to $\sim$ 2\% of the aperture diameter.

The receiver electronics shown in Figure~\ref{fig:system} are installed in the SPT receiver cabin under the optical bench, together with the SPT-3G electronics. An external computer controls mixer and amplifier bias settings. The LO phased locked loop (PLL) incorporates a 100~MHz crystal reference oscillator that is locked to the maser 10~MHz, to which a 12.3~GHz dielectric resonator oscillator (DRO) is locked for the band 6 LO. This LO is locked to the 6$^{\rm th}$ harmonic of the DRO signal with a 100~MHz offset, again referenced to the 100~MHz crystal. The band 7 LO is locked to a 12.7~GHz DRO at the 9$^{\rm th}$ harmonic. The PLL box has computer-adjustable loop parameters and lock monitoring. The warm IF amplifier box encompasses four chains (two sidebands, two polarizations) of amplifiers and computer-controlled variable attenuators. Since the cabin tilts and rotates with the telescope, the IF signal is transferred via optical fiber link to the stationary SPT control room, where the VLBI backend is installed. The iBOB spectrometer is explained in a later section.

\begin{figure}[t]
   \begin{center}
   \begin{tabular}{c}
   \includegraphics[height=7.5cm]{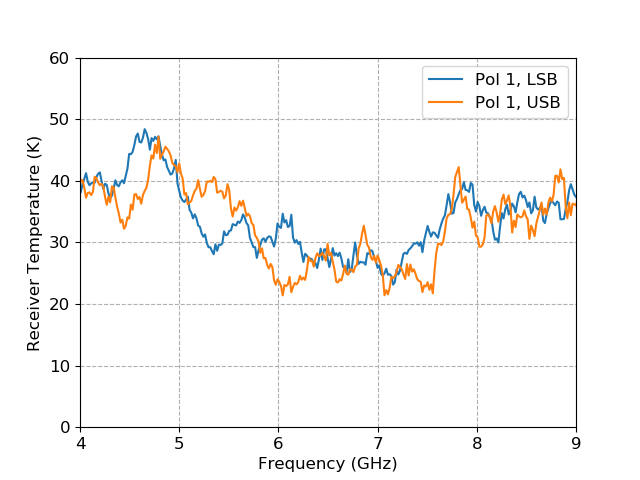}
   \end{tabular}
   \end{center}
   \caption[]
   {\label{fig:Trx} 230~GHz band spectral noise temperature for both sidebands of one polarization. The receiver temperature is $\sim$40~K in the 5-9~GHz IF passband.}
   \end{figure}

We verified the receiver performance with noise temperature measurements, tone injection tests, and the phase stability tests. Figure~\ref{fig:Trx} shows the noise temperature of the band 6 receiver across the IF band for one polarization. It is $\sim$40~K for all four IF channels. This is measured through the Y-factor technique using a nitrogen-temperature cold load. The same test measured at the VLBI backend shows no degradation in noise temperature from the subsequent gain, fiber, and downconversion stages.

\subsection{Frequency reference}

A Hydrogen-maser manufactured by T4Science 
was installed at the DSL to serve as the fundamental frequency reference for VLBI observation. It provides 10 MHz references for synchronizing the receiver system with LO and time-stamping of recorded data. We set up the maser inside a temperature-controlled enclosure placed on a stack of urethane to vibrationally isolate the maser from the DSL building and keep a temperature stable environment. The maser signal is connected to the receiver cabin via a long run of coax: approximately 100 feet of LMR-400 through the interior of the SPT/DSL building and approximately 200 feet of Times Microwave Phase Track 210 (PT210) cable through the azimuth and elevation cable wraps of the SPT. These cables are selected to be particularly phase stable at 10~MHz at their operating temperatures\cite{rogers2008_mk5_069}. The PT210 cable is run through silicone foam sponge tubing to slow thermal changes, and this is further encased in a flexible metal conduit. Both the LMR-400 and PT210 segments are composed of a pair of cables so that the round-trip phase on the 10~MHz can be monitored at the maser. A low noise distribution amplifier (LNDA) inside the receiver cabin distributes the \mbox{10 MHz} coming from the maser and provides a 10~MHz loopback signal that returns on the second coax. During observations the 10~MHz round-trip phase is continuously monitored for changes induced by thermal, mechanical, or other disturbances. The Allan deviation of the round-trip maser phase is virtually identical to what was found by the manufacturer when beating the maser against an identical model, around $1\times10^{-13}$ at 1 s and $1.2\times10^{-15}$ at 1000 s. To ensure that the maser is operating properly before observations, it is compared against an Oscilloquartz crystal oscillator to verify the frequency stability and the phase noise characteristics.

\subsection{VLBI backend}
\label{sec:rxsystem_vlbi}

\begin{figure}[hb]
   \vspace{10pt}
   \begin{center}
   \begin{tabular}{c}
   \includegraphics[height=12cm]{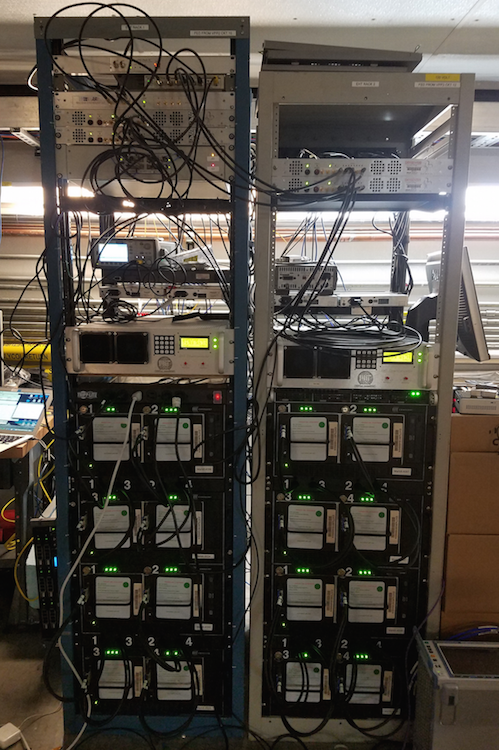}
   \end{tabular}
   \end{center}
   \caption[]
   {\label{fig:vlbirack} VLBI backend installed at the SPT control room. These racks comprise a 64Gbps data recording system.}
   \end{figure}

The VLBI backend is designed to ingest four IF bands from the receiver (two sidebands of two polarizations), each spanning 4~GHz. For band 6 these are 5-9~GHz, for band 7 they are 4-8~GHz. 
The block downconverter (BDC) divides each IF into two 0--2 GHz basebands, using an internal LO at 7 or 6 GHz. When all of these bands are digitized to two bits of precision at the Nyquist rate, the instantaneous recording rate is 64~Gbits per second. The digitization is done by ROACH2 (Reconfigurable Open Architecture Computing Hardware) digital backend (R2DBE) units, which have demonstrated 4096 megasamples per second sampling for two channels\cite{2015PASP..127.1226V}. The 64~Gbps EHT backend system consists of four R2DBEs and four Mark 6 recorders\cite{2011evga.conf...31W, 2013PASP..125..196W}. Figure~\ref{fig:vlbirack} shows the VLBI backend setup at the SPT. The recorded data are correlated on the DiFX correlators \cite{2011PASP..123..275D} at the MIT Haystack Observatory and the Max Planck Institute for Radio Astronomy.

\subsection{Calibration system and spectrometer}
\label{sec:rxsystem_cal}

\begin{figure}[t]
   \begin{center}
   \begin{tabular}{cc}
   \includegraphics[height=7.5cm]{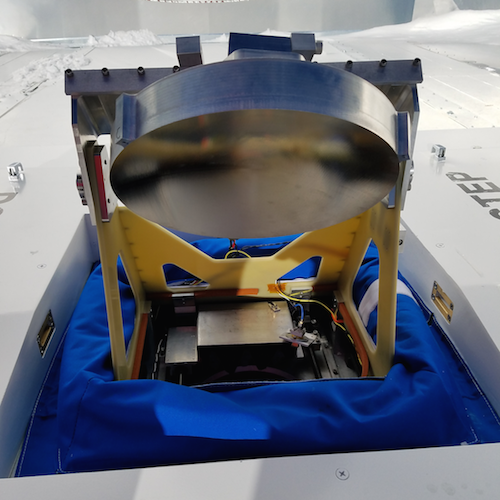} & \includegraphics[height=7.5cm]{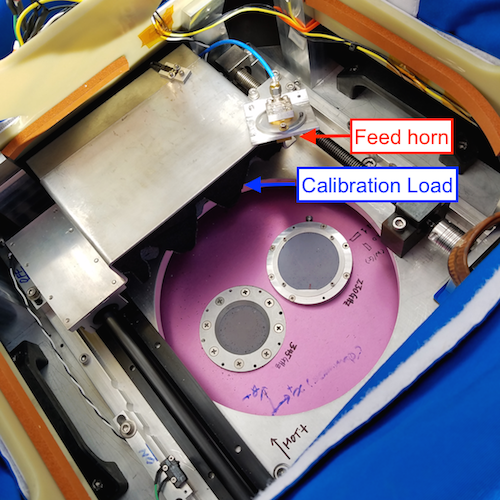}
   \end{tabular}
   \end{center}
   \caption[]
   { \label{fig:cal_tert} {\it Left}: The tertiary mirror assembly installed on top of the receiver dewar for the observation. {\it Right}: The calibration load for the system temperature measurement, on the tertiary mount. A feed horn is located on the calibration load with a harmonic mixer for a tone injection. The tone injection feed horn assembly is tilted downward so that it can illuminate the 230 GHz receiver feed horn.}
   \end{figure}

We developed a calibration system\footnote{Receiver Selection and Calibration Unit for EHT-SPT (RESCUES); http://hdl.handle.net/10150/579318} for the receiver to keep track of the system temperature during the observation. The system temperature can be derived by the ratio of powers received from the sky and a load of known temperature. Due to the limited space inside the receiver cabin, the calibration system is installed above the receiver cryostat, along with the tertiary mirror mount (Figure~\ref{fig:cal_tert}). We use microwave absorber as an ambient load, coupled to a Schneider Electric Motion LMDCE421 motor with a rotation shaft. It covers the receiver beam as we run the calibration. The AD-590 temperature transducer is buried inside the absorber to read the load temperature. The atmospheric opacity is useful information for the calibration because the system temperature depends on a source elevation. At the telescope site, we have a 350~$\mu$m tipping radiometer installed and can convert 350~$\mu$m opacity to 225 GHz opacity by a conversion relation given in Ref.~\citenum{2016PASP..128g5001R}. The calibration load also carries a feed horn and a harmonic mixer that can be positioned in the receiver beam to generate a coherent tone for signal path verification and coherence testing.

\begin{figure}[t]
   \begin{center}
   \begin{tabular}{c}
   \includegraphics[height=10cm]{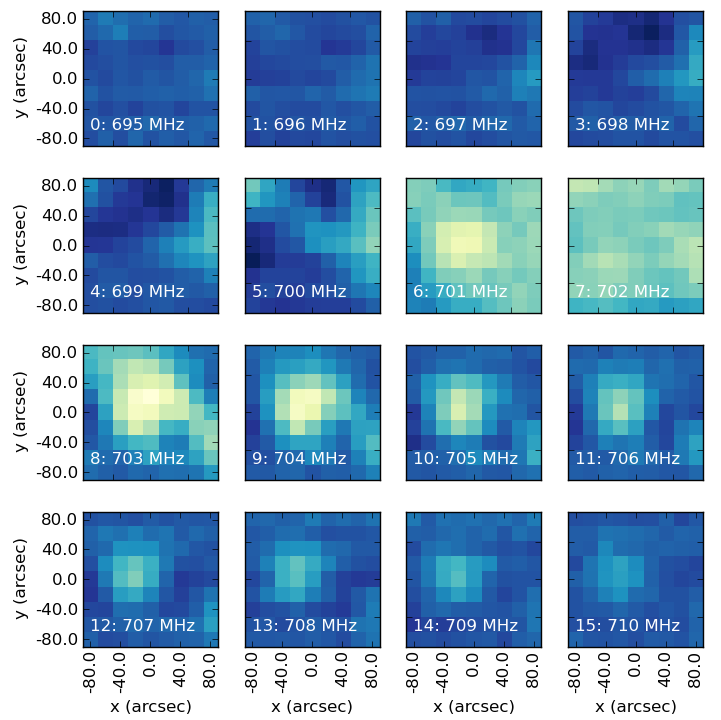}
   \end{tabular}
   \end{center}
   \caption[]
   {\label{fig:ibob_spec} Spectral channel map of the protostar IRAS 16293-2422, a pointing source. The map shows 16 1~MHz channels starting from a second IF of 695 MHz. Pointing offsets can be determined from one or many channels, as appropriate for each source.}
   \end{figure}

To aid in pointing, we installed a digital spectrometer for measurements of CO lines that lie near to the EHT observing bands. The spectrometer itself is a pair of FPGA spectrometers based on the CASPER iBOB board\footnote{https://casper.berkeley.edu/wiki/1\_GHz\_-\_1024\_Channel\_Wideband\_Spectrometer}. They are fed by a special signal chain that converts the CO 2-1 line (230.538 GHz, IF=9.438 GHz for band 6) and CO 3-2 (345.796 GHz, IF=3.196 GHz for band 7) to approximately 750~MHz. An example of a spectral line map with this system is provided in Figure~\ref{fig:ibob_spec}. 


\section{OPTICS}
\label{sec:optics}

The SPT has an off-axis Gregorian design with a 10-meter primary mirror to minimize blockage and scattering of incident light from the faint CMB (see Ref.~\citenum{2008ApOpt..47.4418P} for detail). The SPT was not initially designed to illuminate any instrument other than its CMB camera, so special optics are required to redirect the light from the primary mirror to the VLBI receiver. 
\subsection{Design}
\label{sec:optics_design}

\begin{figure}[ht]
   \begin{center}
   \begin{tabular}{c}
   \includegraphics[height=7.5cm]{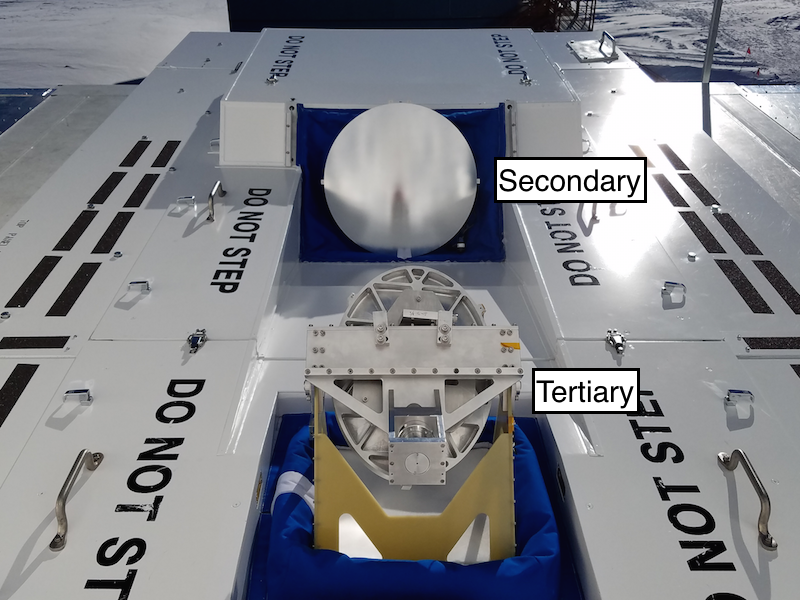}
   \end{tabular}
   \end{center}
   \caption[]
   {\label{fig:vlbioptics} The secondary and tertiary mirrors of the SPT VLBI receiver system installed at the South Pole, viewed from the primary mirror. The mirror assemblies are covered by environmental seals to prevent cold air flowing into the receiver cabin.}
   \end{figure}

The VLBI optical system has a Cassegrain design, with hyperbolic secondary and ellipsoidal tertiary mirrors. The mirror parameters were optimized with the Zemax optical design software. The model was chosen such that the optics illuminate the 10 m dish to greater than 12 dB at both frequencies, given the beam parameters and locations of the 230 and 345 GHz feed horns. In Figure~\ref{fig:optics_model}, we show the VLBI optics installed around the SPT-3G receiver and its optics. The VLBI secondary mirror blocks the 3G secondary mirror and reflects the beam from the primary to the VLBI tertiary mirror, and then to feed horns of the VLBI receiver. The tertiary mirror is mounted to the top of the dewar and, to simplify winter operation, this mirror rotates around the optical axis so that it focuses the beam towards either 230 or 345 GHz side of the receiver. 
The mirrors are easily removable to clear the optical path for the CMB receiver and only installed for the EHT observing campaigns. We use a SpitzLift portable crane for transport of the mirrors to the receiver cabin top. Figure~\ref{fig:vlbioptics} shows the secondary and tertiary mirrors installed for VLBI observation.

\subsection{Beam measurement}
\label{sec:optics_beam}

\begin{figure}[ht]
   \begin{center}
   \begin{tabular}{cc}
   \includegraphics[height=6cm]{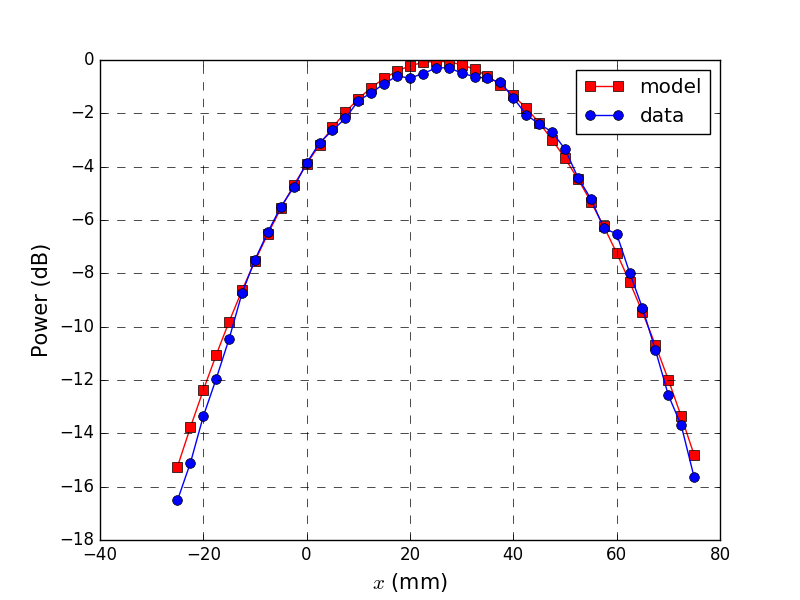} & \includegraphics[height=6cm]{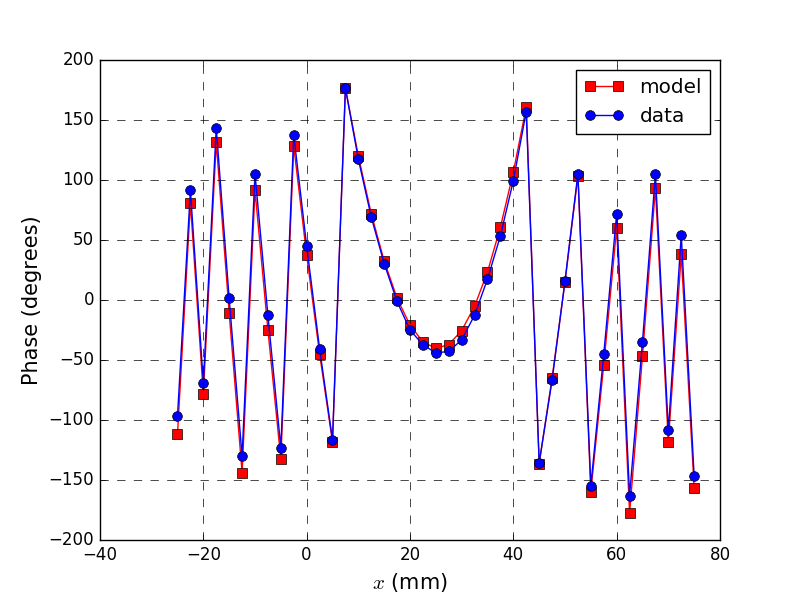}
   \end{tabular}
   \end{center}
   \caption[] 
   { \label{fig:rx_beam} The vector beam measurement of the 230 GHz receiver provides both power ({\it left}) and phase ({\it right}). The blue dots and the red squares show the data and the best fit model.}
   \end{figure}

The receiver beam pattern has been measured in both frequency bands (Figure~\ref{fig:rx_beam}) by measuring the response to a coherent tone as it is scanned across a plane above the receiver\cite{2018KimMarroneISSTT}. As shown in Figure~\ref{fig:rx_assy}, both 230 and 345 GHz receivers are inside a single dewar and the feed horns are intentionally tilted inward, toward the centerline of the receiver between the two horns, so that they both can face the shared tertiary. We model the near-field scan as a Gaussian beam propagating at an angle to the measurement plane. We characterize the model parameters, including the three-dimensional location of the feed horn phase center and its tilt angle, using the fit between the model and the data. The inferred parameters indicate that the feed horn assemblies are correctly positioned and oriented.

We also measured the beam pattern after the tertiary mirror using the same technique in planes near the Cassegrain focus. These data were used to verify the beam propagation direction between tertiary and secondary, based on the measured propagation angle and the position of the beam in several parallel planes along the propagation direction.


\section{SOFTWARE}
\label{sec:software}

\begin{figure}[t]
   \begin{center}
   \begin{tabular}{c}
   \includegraphics[height=8cm]{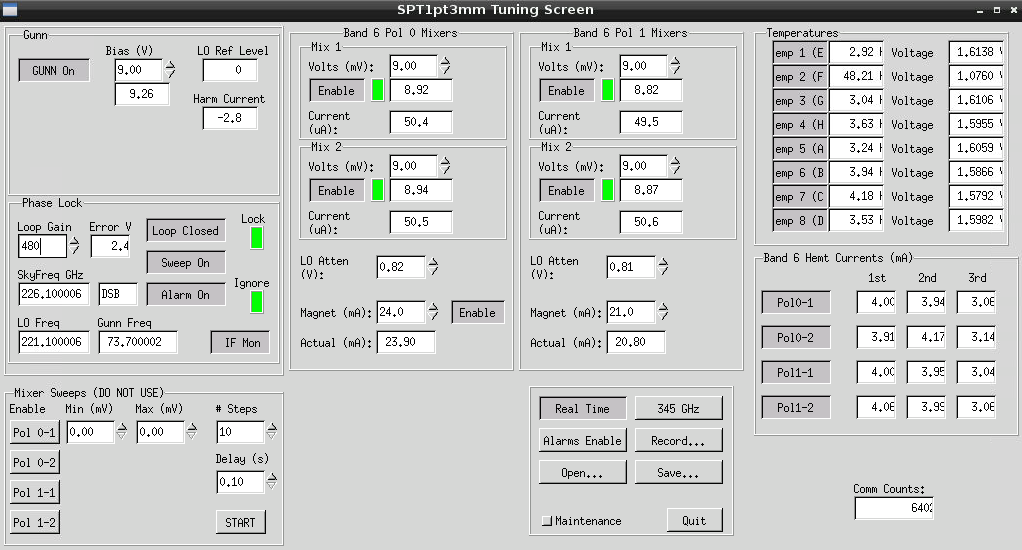}
   \end{tabular}
   \end{center}
   \caption[]
   {\label{fig:tuningscreen} The tuning screen for the 230 GHz receiver. The screen controls mixer bias, the Gunn oscillator lock, and LO injection, and monitors the temperature inside the receiver.}
   \end{figure}

The SPT VLBI receiving system has three types of software: receiver control, calibration, and telescope control. The receiver control software runs on a BeagleBone Black (BBB), a single-board computer that supports the Linux environment. The software controls the receiver and related electronics until the optical fiber relay. The receiver tuning screen in Figure~\ref{fig:tuningscreen} is the primary control interface and was originally developed by Thomas W. Folkers for the Kitt Peak 12-m and Submillimeter Telescope (SMT) on Mount Graham, operated by the Arizona Radio Observatory (ARO). This software has been adapted for the SPT application. 
The primary parameters controlled/monitored by this software are the mixer and amplifier bias settings, local oscillator power levels and phase-locked loop parameters, thermometry, and the gain of the warm amplifier chain.

The calibration software obtains data for the system temperature measurement and a priori calibration of the telescope. It involves the monitoring of the calibration load temperature and the IF power, and positioning of feed horn for the tone injection. The software interacts with the telescope control software, Generic Control Program (GCP)\cite{Story:2012dr}, by which the SPT control and data acquisition are done.

\section{SUMMARY}

In this paper, we describe the development of the VLBI receiver for the SPT. We deployed the receiver system, including a hydrogen maser and VLBI recording backend system, to the South Pole in the 2016-17 austral summer, and the SPT  joined the EHT array in its first campaign in April 2017. This system samples two polarizations near LO frequencies of either 221.1 or 342.6 GHz and instantaneously digitizes 16~GHz of receiver bandwidth, yielding a 64~Gbps VLBI data rate. The clean receiver optical path, low-noise mixers, and the atmospheric environment of the South Pole combine to create a high-sensitivity VLBI station that significantly extends the baseline coverage of the EHT array.

\acknowledgments 
J.K. and D.P.M. acknowledge support from NSF grants AST-1207752 and AST-1440254. The South Pole Telescope program is supported by the National Science Foundation through grant PLR-1248097. Partial support is also provided by the NSF Physics Frontier Center grant PHY-0114422 to the Kavli Institute of Cosmological Physics at the University of Chicago, the Kavli Foundation, and the Gordon and Betty Moore Foundation through Grant GBMF\#947 to the University of Chicago. We thank Chris Kendall and Dave Pernic for their assistance at the South Pole. We acknowledge essential support for this system that is provided through the loan of several key components. The maser and quartz crystal are on loan from the Academia Sinica Institute of Astronomy and Astrophysics. The National Radio Astronomy Observatory has loaned the band 6 feed horn, which does not meet ALMA specifications at non-EHT frequencies. The Smithsonian Astrophysical Observatory has loaned cryogenic IF amplifiers that are used in the band 7 receiving system.

\bibliography{sptvlbirx}
\bibliographystyle{spiebib} 

\end{document}